\documentclass
[superscriptaddress,secnumarabic,amssymb,amsmath,nobibnotes,aps,prd,nofootinbib,column,notitlepage,10pt]{revtex4}%
\usepackage{setspace}
\usepackage{color}
\usepackage{amsmath}
\usepackage{amsfonts}
\usepackage{verbatim}
\usepackage{amssymb}
\usepackage[colorlinks]{hyperref}
\usepackage{graphicx,bm}
\usepackage{amsmath}
\usepackage{amssymb}
\usepackage{graphicx}
\usepackage{textgreek}
\usepackage[caption=false]{subfig}%
\providecommand{\U}[1]{\protect\rule{.1in}{.1in}}

\newcommand{\be}{\begin{equation}}
\newcommand{\ee}{\end{equation}}

\newcommand{\mincir}{\raise
-3.truept\hbox{\rlap{\hbox{$\sim$}}\raise4.truept\hbox{$<$}\ }}
\newcommand{\magcir}{\raise
-3.truept\hbox{\rlap{\hbox{$\sim$}}\raise4.truept\hbox{$>$}\ }}

\ifx\pdfoutput\relax\let\pdfoutput=\undefined\fi
\newcount\msipdfoutput
\ifx\pdfoutput\undefined\else
\ifcase\pdfoutput\else
\ifx\paperwidth\undefined\else
\ifdim\paperheight=0pt\relax\else\pdfpageheight\paperheight\fi
\ifdim\paperwidth=0pt\relax\else\pdfpagewidth\paperwidth\fi
\fi\fi\fi
\begin{document}
\title{Late-Time Cosmological Constraints on Kaniadakis Holographic  Dark Energy}
\author{Giuseppe Gaetano Luciano}
\email{giuseppegaetano.luciano@udl.cat}
\affiliation{Departamento de Qu\'{\i}mica, F\'{\i}sica y Ciencias Ambientales y del Suelo,
Escuela Polit\'{e}cnica Superior -- Lleida, Universidad de Lleida, Av. Jaume
II, 69, 25001 Lleida, Spain}
\author{Andronikos Paliathanasis}
\email{anpaliat@phys.uoa.gr}
\affiliation{Department of Mathematics, Faculty of Applied Sciences, Durban University of
Technology, Durban 4000, South Africa}
\affiliation{National Institute for Theoretical and Computational Sciences (NITheCS), South Africa}
\affiliation{Departamento de Matem\`{a}ticas, Universidad Cat\`{o}lica del Norte, Avda.
Angamos 0610, Casilla 1280 Antofagasta, Chile}

\begin{abstract}
Kaniadakis Holographic Dark Energy is a one-parameter extension of the standard HDE framework, in which the horizon entropy is reformulated using Kaniadakis entropy. 
At the cosmological level, it has been shown to give rise to modified Friedmann equations, leading to a richer phenomenology compared to $\Lambda$CDM.  In this work we test the Kanadiakis holography model against multiple late-time observational probes, including Type Ia supernovae from PantheonPlus and Union3, Cosmic Chronometer measurements of the Hubble parameter and Baryon Acoustic Oscillation from the Dark Energy Spectroscopic Instrument Data Release 2 (DESI DR2). Using a Bayesian inference approach with MCMC methods, we constrain the cosmological parameters of the model and evaluate its performance against $\Lambda$CDM through the Akaike Information Criterion. We find that Kaniadakis holography can provide a better fit for some data combinations, although $\Lambda$CDM remains slightly statistically favored overall. These results highlight Kaniadakis holography as a competitive alternative to the standard cosmological model, offering valuable insights into the role of generalized entropy in dark energy dynamics.


\end{abstract}

\date{\today}
\maketitle

\section{Introduction}

\label{sec1}

\textit{Holographic dark energy} (HDE) has emerged as a compelling 
alternative framework for addressing the long-standing dark energy problem in cosmology. 
Unlike conventional approaches, it is founded on the \textit{holographic principle}, 
a fundamental idea in quantum gravity which suggests that the description of a physical 
system within a given volume can be fully encoded on its boundary surface 
\cite{tHooft:1993dmi,Susskind:1994vu,Bousso:2002ju}. 

Within this paradigm, the connection between ultraviolet (UV) and infrared (IR) cutoffs 
in an effective quantum field theory plays a central role. In particular, it has been 
argued that the total vacuum energy of a system must not exceed the energy of a black 
hole of the same size, thereby linking the UV cutoff (associated with short-distance 
quantum fluctuations) to a maximal IR scale (related to the largest observable length) 
\cite{Cohen:1998zx}.  Interpreted in a cosmological context, this restriction leads to a holographic estimate 
of the vacuum energy density, which can be naturally associated with the dark energy 
responsible for the accelerated expansion of the universe \cite{Li:2004rb,Wang:2016och}. 
Thus, the HDE framework provides a bridge between principles of quantum gravity and 
large-scale cosmological dynamics, offering both theoretical motivation and 
phenomenological relevance. 

A central issue in implementing the holographic principle in cosmology lies in 
specifying a physically consistent IR cutoff and understanding its thermodynamic 
implications. A key insight is that the entropy of a cosmological horizon scales 
with its surface area rather than its volume, in analogy with the 
Bekenstein--Hawking entropy of black holes 
\cite{Bekenstein:1973ur,Hawking:1975vcx}. Building on this idea, the original HDE 
model identified the future event horizon as the IR cutoff and imposed the 
Bekenstein--Hawking area law as the entropy bound. This formulation produces a 
dynamical dark energy scenario consistent with a wide range of observational 
constraints, including type Ia supernovae, cosmic microwave background anisotropies, 
and baryon acoustic oscillations 
\cite{Zhang:2005hs,Li:2009bn,Feng:2007wn,Zhang:2009un,Lu:2009iv}.  

Nevertheless, the original scenario admits a variety of extensions. Early 
developments explored the consequences of adopting alternative IR cutoffs as well 
as possible interactions between the dark energy and dark matter sectors 
\cite{Wang:2016och,Wang:2016lxa}. More recently, inspired by generalized 
statistical mechanics, a number of modified HDE models have been formulated using 
non-standard entropy definitions. Notable examples include the frameworks based on 
Tsallis entropy \cite{Tsallis:1987eu,Tsallis:2013}, Kaniadakis entropy 
\cite{kaniadakis2001non}, R\'enyi entropy \cite{renyi1961entropy}, and Barrow 
entropy \cite{Barrow:2020tzx}, among others. These generalized models and their 
cosmological consequences have been extensively investigated in the literature 
\cite{Horvat:2004vn,Huang:2004ai,Pavon:2005yx,
Wang:2005jx,Nojiri:2005pu,Wang:2005ph,Setare:2006wh,Setare:2008pc,
Gong:2004fq,Saridakis:2007cy,Setare:2007we,Cai:2007us,Saridakis:2007ns, 
Jamil:2009sq,Micheletti:2009jy,Aviles:2011sfa,Chimento:2011pk,Pourhassan:2017cba,
Tavayef:2018xwx,Saridakis:2018unr,Nojiri:2019kkp,Geng:2019shx,DAgostino:2019wko,
Saridakis:2020zol,Drepanou:2021jiv,Luciano:2022hhy,Luciano:2022ffn,
Nakarachinda:2023jko,Mamon:2020spa,Ghaffari:2022skp,Luciano:2023roh,Luciano:2023wtx, Lambiase:2023ryq,Capozziello:2025axh}.  

Notably, the Kaniadakis entropy has attracted particular attention in recent 
years \cite{Luciano:2022eio}. Originally proposed within the framework of relativistic statistical mechanics, 
it introduces a (dimensionless) deformation parameter $K$ that smoothly interpolates between 
classical Boltzmann--Gibbs statistics and non-extensive generalizations 
\cite{kaniadakis2001non}. The resulting entropy takes the form
\begin{equation}
\label{Ken}
S_K=-k_B\sum_{i=1}^{W}\frac{P_i^{1+K}-P_i^{1-K}}{2K}\,,
\end{equation}
where \(P_i\) is the probability that the system occupies microstate \(i\) and \(W\) is the total number of configurations. It is straightforward to verify that the standard entropy is recovered in the limit \(K \to 0\), while \(K\) is restricted to the range \(-1 < K < 1\).

When applied to black–hole thermodynamics - which is of direct relevance for holographic
analysis - the Kaniadakis entropy can be formulated under the equiprobability
assumption \(P_i = 1/W\). Since the Boltzmann–Gibbs entropy satisfies \(S \propto \ln W\)
and, for a black–hole horizon, the Bekenstein–Hawking entropy is $S_{{BH}} = \frac{A}{4G}$, the total number of configurations follows as $W = \exp\!\left(\frac{A}{4G}\right)$, 
where, unless stated otherwise, we work 
in units with the Boltzmann constant, the speed of light, and Planck’s constant 
set to $k_{B} = c = \hbar = 1$.
Substituting this expression for \(W\) into Eq.~\eqref{Ken} then yields 
\begin{equation}
\label{KE}
    S_K=\frac{1}{K}\sinh\left(K S_{BH}\right).
\end{equation}
As anticipated, in the limit \(K \to 0\) the Kaniadakis entropy reduces to the
Bekenstein–Hawking result, $\lim_{K\to 0} S_{K} = S_{BH}$. Since in realistic situations the modified entropy is expected to differ only
slightly from the standard value, one can take \(|K|\ll 1\). Expanding for small \(K\) then gives
\begin{equation}
\label{Kapp}
S_{K} = S_{{BH}} + \frac{K^{2}}{6}\,S_{{BH}}^{3}
+ \mathcal{O}\!\left(K^{4}\right).
\end{equation}
The first term reproduces the usual Bekenstein–Hawking entropy, while the second
term is the leading Kaniadakis correction.

The Kaniadakis-entropy extension of the HDE model has been investigated in \cite{Drepanou:2021jiv,Hernandez-Almada:2021aiw,Sharma:2021zjx}, where it was shown to reproduce the Universe’s thermal history, including the standard succession of matter- and dark-energy–dominated eras. Interestingly, the entropic parameter plays a pivotal role in shaping the dark-energy equation of state, enabling quintessence-like behavior, entry into the phantom regime, or even crossings of the phantom divide over cosmic time.

From an observational standpoint, 
the Second Data Release of the Dark Energy Spectroscopic Instrument (DESI~DR2) survey \cite{des4,des5,des6} has already proven invaluable for imposing stringent constraints across a broad range of cosmological models. Thanks to the high precision of baryon acoustic oscillation (BAO) measurements, the DESI data set enables stringent tests of numerous extensions of the standard \(\Lambda\)CDM framework. It has been employed to constrain dynamical dark energy models~\cite{Ormondroyd:2025iaf,You:2025uon,Gu:2025xie,Santos:2025wiv,Alfano:2025gie,Carloni:2024zpl,Luciano:2025elo,Luciano:2025hjn}, early dark energy scenarios~\cite{Chaussidon:2025npr}, and broad classes of scalar-field theories with both minimal and non-minimal couplings~\cite{Anchordoqui:2025fgz,Ye:2025ulq,Wolf:2025jed,Samanta:2025oqz,Alfano:2025awf}. BAO data have also been used to probe quantum-gravity-inspired frameworks, such as those based on the Generalized Uncertainty Principle~\cite{Paliathanasis:2025dcr,Paliathanasis:2025kmg}, as well as interacting dark-sector models~\cite{vanderWesthuizen:2025iam,Shah:2025ayl,Silva:2025hxw,Pan:2025qwy,Petri:2025swg}. Additional applications include astrophysical tests~\cite{Alfano:2024jqn}, model-independent cosmographic reconstructions~\cite{Luongo:2024fww}, and a wide range of modified-gravity and modified-entropy theories~\cite{Yang:2025mws,Li:2025cxn,Paliathanasis:2025hjw,Tyagi:2025zov}, along with various other scenarios~\cite{Brandenberger:2025hof,Paliathanasis:2025cuc,Ishiyama:2025bbd,Wang:2025ljj,Akrami:2025zlb,Colgain:2025nzf,Plaza:2025gjf,Toda:2025dzd,Dinda:2025svh,Kumar:2025etf,Mirpoorian:2025rfp,deSouza:2025rhv,Scherer:2025esj,Preston:2025tyl,Abedin:2025yru,Wang:2025vfb,Bayat:2025xfr,Cai:2025mas,Ye:2025ark,Andriot:2025los,RoyChoudhury:2025dhe,Cheng:2025lod,Fazzari:2025lzd,Zhou:2025nkb,Li:2025dwz,Karmakar:2025iba,Gialamas:2025pwv}.

Starting from the above premises, 
in this work we use observational data from Type Ia supernovae ({PantheonPlus} and {Union3}), direct \(H(z)\) measurements from cosmic chronometers (CC), and BAO from DESI~DR2 to constrain the Kaniadakis deformation parameter, which quantifies departures from the standard entropy–area relation.
The structure of the paper is as follows: in Sec. \ref{sec2} we present the Kaniadakis Holographic Dark Energy (KHDE) scenario. Observational analysis is conducted in Sec. \ref{sec3}, while conclusions and outlook are summarized in Sec. \ref{sec4}.

\section{Kaniadakis Holographic Dark Energy}
\label{sec2}

Following \cite{Drepanou:2021jiv}, in this section we formulate the generalized HDE framework starting from Eq. \eqref{Kapp}. Toward this end, we recall that in the conventional HDE model the constraint on the dark energy density can be expressed through the condition \( \rho_{{DE}} L^4 \leq S \), where \( L \) denotes the infrared (IR) cutoff scale. If the entropy obeys the Bekenstein–Hawking area law, scaling as \( S \propto A \propto L^2 \)~\cite{Wang:2016och}, one immediately recovers the standard HDE scenario. 

By contrast, adopting the Kaniadakis entropy modification \eqref{Kapp} leads to  
\begin{equation}
\label{rde}
   \rho_{{DE}} = 3c^2m_P^2 L^{-2}+3\tilde c^2K^2m_P^6L^2\,,
\end{equation}
where $c,\tilde c$ are suitable constants and $m_P=(8\pi G)^{-1/2}$ is the reduced Planck mass. As previously noted, for \( K = 0 \) the above relation reduces to the standard HDE expression, namely \( \rho_{{DE}} = 3c^{2}m_{p}^{2}L^{-2} \). Following \cite{Hernandez-Almada:2021aiw}, in our next analysis we shall set $3\tilde{c} \sim \mathcal{O}(1)$. This assumption is reasonable as long as the constant $\tilde{c}$ is comparable to $c$, which, according to many observational analyses, is estimated to be of order unity \cite{Li:2013dha}.

To explore the consequences of these non-standard evolutionary features, we assume a spatially flat, homogeneous and isotropic Universe, characterized by the Friedmann–Robertson–Walker (FRW) metric, 
\begin{equation}
ds^2 = -dt^2+a^2(t)\delta_{ij}dx^idx^j
\, ,
\label{FRW}
\end{equation}
where  $a(t)$ is the time-dependent scale factor.

As a subsequent step, within any HDE framework it is essential to specify the length scale \( L \). In the context of conventional HDE models, it is well established that identifying \( L \) with the Hubble horizon \( H^{-1} \) (with \( H \equiv \dot{a}/a \) the Hubble parameter) is not viable, since this choice leads to well-known inconsistencies~\cite{Hsu:2004ri}, such as the absence of cosmic acceleration. Consequently, the future event horizon is typically adopted\footnote{For a more detailed discussion on the advantages of employing the future horizon as the IR cutoff, its implications in KHDE, and the distinctions from alternative models based on different cutoff choices, see \cite{Drepanou:2021jiv}.}\cite{Li:2004rb}, 
\begin{equation}
    \label{IRc}
R_h = a \int_t^{\infty} \frac{dt'}{a(t')} = a \int_a^\infty \frac{da'}{Ha'^2} \, .
\end{equation}
Therefore, the energy density \eqref{rde} of KHDE takes the form
\begin{equation}
\label{rKHDE}
       \rho_{{DE}} = 3c^2m_P^2 R_h^{-2}+K^2m_P^6R_h^2\,,
\end{equation}

Let us assume the Universe to contain the usual matter sector, described as a perfect fluid with energy density $\rho_m$, together with the KHDE component introduced earlier. Within this framework, the Friedmann equations take the form  
\begin{eqnarray}
\label{FFE}
\rho_m + \rho_{{DE}} &=& 3m_p^2 H^2\,, \\[2mm]
\rho_m + p_m + \rho_{{DE}} + p_{{DE}} &=& -2m_p^2 \dot H\,.
\label{SFE}
\end{eqnarray}
where $p_m$ and $p_{{DE}}$ denote the pressures of matter and KHDE, respectively. Defining the dimensionless density parameters as  
\begin{equation}
\label{frde}
\Omega_m \equiv \frac{\rho_m}{3m_p^2 H^2}\,, 
\qquad 
\Omega_{{DE}} \equiv \frac{\rho_{{DE}}}{3m_p^2 H^2}\,,
\end{equation}
Eq.~\eqref{FFE} can be recast in the compact form  $\Omega_m + \Omega_{{DE}} = 1$. A further constraint follows from the matter continuity equation
\begin{equation}
\label{contmat}
\dot{\rho}_m + 3H\left(\rho_m + p_m\right) = 0\,.
\end{equation}
From Eqs. \eqref{IRc}, \eqref{rKHDE} and \eqref{frde}, we then get \cite{Drepanou:2021jiv}
\begin{equation}
    \int_x^{\infty}\frac{dx}{Ha}=\frac{1}{a}\left(\frac{3H^2\Omega_{DE}-\sqrt{9H^4\Omega^2_{DE}-12c^2K^2m_P^4}}
    {2K^2m_P^4}
    \right)^{\frac{1}{2}}\,,
    \label{int}
\end{equation}
where we have used the notation $x\equiv \log a$. 

We now examine the case of pressureless matter, for which 
$\rho_{m} = \rho_{m0}/a^{3}$ and 
$\Omega_{m} = \Omega_{m0} H_{0}^{2}/(a^{3} H^{2})$. 
Here we set $a_{0} = 1$, and the subscript ``0'' denotes the present value of the corresponding quantity.
Using the compact form of the Friedmann equation \eqref{FFE}, we find the following expression for 
the Hubble function  
\begin{equation}
\label{H}
H=\frac{H_{0}\sqrt{\Omega_{m0}}}{\sqrt{a^{3}\left(  1-\Omega_{DE}\right)}}.
\end{equation}
By inserting this equation into \eqref{int}, after some manipulation we obtain the following differential equation for the KHDE component:
\begin{equation}
\label{diffom}
\frac{d\Omega_{DE}}{dx}=\Omega_{DE}\left(  1-\Omega_{DE}\right)  \left\{
3-\frac{2\left(\mathcal{A}-2 K^2 m_p^4 \mathcal{B}\right)}{\mathcal{A}}\left[  1-\left( 3\frac{\Omega_{DE}}%
{\mathcal{AB}}\right)  ^{\frac{1}{2}}\right]  \right\},
\end{equation}
where we have defined
\begin{eqnarray}
\mathcal{A}\left(  H_{0},\Omega_{m0},x,\Omega_{DE}\right)   &  =&\frac{3H_{0}^{2}%
\Omega_{m0}\Omega_{DE}}{\left(  1-\Omega_{DE}\right) e^{3x}}\,,\\
\mathcal{B}\left(  H_{0},\Omega_{m0},x,\Omega_{DE},K^2 m_p^4,c\right)&=&
\frac{\mathcal{A}\left(
H_{0},\Omega_{m0},x,\Omega_{DE}\right)  -\sqrt{\mathcal{A}^2\left(H_{0},\Omega
_{m0},x,\Omega_{DE}\right)-12K^2 m_p^4 c^2}}{2 K^2 m_p^4}\,.%
\end{eqnarray}
Finally, in terms of the redshift $z=1/a-1$, Eqs. \eqref{H} and \eqref{diffom} read
\begin{eqnarray}
H(z)  &=&\frac{H_{0}\sqrt{\Omega_{m0}\left(  1+z\right)  ^{3}}%
}{\sqrt{\left(  1-\Omega_{DE}\right)  }}\,,\\
-\left(  1+z\right)  \frac{d\Omega_{DE}}{dz}&=&\Omega_{DE}\left(  1-\Omega_{DE}\right)  \left\{
3-\frac{2\left(\mathcal{A}-2 K^2 m_p^4 \mathcal{B}\right)}{\mathcal{A}}\left[  1-\left( 3\frac{\Omega_{DE}}%
{\mathcal{AB}}\right)  ^{\frac{1}{2}}\right]  \right\}\,.
\end{eqnarray}
These equations determine the evolution of the Hubble parameter and HDE in Kaniadakis entropy-based framework.

\section{Observational Data Analysis}
\label{sec3}
In this section, we employ observational data sets to derive constraints on the parameters of KHDE. 

\subsection{Observational Data}
We begin by outlining the data utilized in our analysis: 

\begin{itemize}
\item Supernova PantheonPlus (PP): This catalogue comprises 1701 light curves from 1550 spectroscopically confirmed supernova events. The data set provides measurements of the distance modulus $\mu^{\mathrm{obs}}$ at redshifts in the range $10^{-3}<z<2.27$~\cite{pan}. The theoretical distance modulus is defined as $\mu^{\mathrm{th}} = 5\log_{10}D_{L} + 25$, where, in a spatially flat FLRW geometry, the luminosity distance is expressed in terms of the Hubble function as $D_{L}(z) = (1+z)\int_{0}^{z}dz^{\prime}/H(z^{\prime})$. In our analysis, we use the PantheonPlus catalogue without applying the SH0ES Cepheid calibration.

\item Supernova of Union3 (U3): This is the most recent supernova catalogue, 
including 2,087 events within the same redshift range as the PP data, of which 
1,363 are shared with the PantheonPlus catalogue~\cite{union}.

\item Observational Hubble Data (OHD): We use direct measurements of the
Hubble parameter obtained from Cosmic Chronometers (CC). These
observations are model-independent, as they do not rely on any
cosmological assumptions. Cosmic Chronometers are passively evolving
galaxies with synchronous stellar populations and similar cosmic
evolution~\cite{co01}. In this analysis, we use 31 direct measurements
of the Hubble parameter in the redshift range $0.09 \leq z \leq 1.965$
as reported in~\cite{cc1}.

\item Baryonic Acoustic Oscillations (BAO): We consider the recent release of
the Dark Energy Spectroscopic Instrument (DESI DR2) BAO observations~\cite{des4,des5,des6}.
This data set provides measurements of the transverse comoving angular distance ratio
$\frac{D_{M}}{r_{drag}}=\frac{D_{L}}{(1+z)\,r_{drag}},$
the volume-averaged distance ratio
$\frac{D_{V}}{r_{drag}}=\frac{(zD_{H}D_{M}^{2})^{1/3}}{r_{drag}},$
and the Hubble distance ratio
$\frac{D_{H}}{r_{drag}}=\frac{1}{r_{drag}H(z)},$
at seven distinct redshifts, where $D_{L}$ is the luminosity distance and
$r_{drag}$ denotes the sound horizon at the drag epoch. In the following analysis,
$r_{drag}$ is treated as a free parameter.
\end{itemize}

%

\begin{table}[t]
\centering
\begin{minipage}{0.48\textwidth}
\centering
\caption{Data sets}
\begin{tabular}{ccccc}\hline\hline
\textbf{Data} & $\mathbf{PP}$ & $\mathbf{U3}$ & $\mathbf{CC}$ & $\mathbf{BAO}$\\\hline
$\mathbf{D}_{1}$ & $\times$ & $\times$ & $\checkmark$ & $\checkmark$\\
$\mathbf{D}_{2}$ & $\checkmark$ & $\times$ & $\times$ & $\checkmark$\\
$\mathbf{D}_{3}$ & $\checkmark$ & $\times$ & $\checkmark$ & $\times$\\
$\mathbf{D}_{4}$ & $\checkmark$ & $\times$ & $\checkmark$ & $\checkmark$\\
$\mathbf{D}_{5}$ & $\times$ & $\checkmark$ & $\times$ & $\checkmark$\\
$\mathbf{D}_{6}$ & $\times$ & $\checkmark$ & $\checkmark$ & $\times$\\
$\mathbf{D}_{7}$ & $\times$ & $\checkmark$ & $\checkmark$ & $\checkmark$\\\hline\hline
\end{tabular}
\label{dataset}
\end{minipage}\hspace{-1cm}
\begin{minipage}{0.48\textwidth}
\centering
\caption{Priors of the Free Parameters}
\begin{tabular}{ccc}\hline\hline
\textbf{Priors} & \textbf{KHDE} & $\Lambda$CDM\\[1mm]\hline
$H_{0}$ & $[60,80]$ & $[60,80]$\\[2.5mm]
$\Omega_{m0}$ & $[0.01,0.4]$ & $[0.01,0.4]$\\[2.5mm]
$K^{2}M_{p}^{4}$ & $[0,1]$ & --\\[2.5mm]
$c$ & $[10^{-3},100]$ & --\\[2.5mm]
$r_{drag}$ & $[130,160]$ & $[130,160]$\\[1mm]\hline\hline
\end{tabular}
\label{prior}
\end{minipage}
\end{table}


\subsection{Methodology}

To carry out the statistical analysis, we employ the Bayesian inference framework 
\textsc{Cobaya}\footnote{\url{https://cobaya.readthedocs.io/}}~\cite{cob1,cob2}, 
using a custom theoretical implementation in combination with the 
MCMC sampler~\cite{mcmc1,mcmc2}. The resulting MCMC chains are analyzed with the 
\textsc{GetDist} library\footnote{\url{https://getdist.readthedocs.io/}}~\cite{getd}.

We consider seven different combinations of data sets, as summarized in Table~\ref{dataset}. 
For each case, we determine the best-fit parameters that maximize the likelihood, 
$\mathcal{L}_{\max} = \exp\!\left(-\tfrac{1}{2}\chi_{\min}^{2}\right)$, where
\begin{equation}
\chi_{\min}^{2} = \chi_{\min(\mathrm{data1})}^{2} + \chi_{\min(\mathrm{data2})}^{2} + \cdots~.
\end{equation}
Furthermore, we apply the same observational tests to the $\Lambda$CDM model. 
Given the different numbers of degrees of freedom in the KHDE and $\Lambda$CDM frameworks, 
we employ the Akaike Information Criterion (AIC)~\cite{AIC} to perform a statistical 
comparison between the two models. The AIC is defined as
\begin{equation}
\mathrm{AIC} \simeq \chi_{\min}^{2} + 2\kappa ,
\end{equation}
where $\kappa$ denotes the number of free parameters of the model.

We adopt Akaike’s scale, which provides a criterion for assessing which model 
offers a better fit to the data, based on the difference 
$\Delta \mathrm{AIC} = \mathrm{AIC}_{\mathrm{KHDE}} - \mathrm{AIC}_{\Lambda}$. 
For the two models under consideration, and noting that the KHDE framework includes two additional free parameters compared to $\Lambda$CDM (see Tab.~\ref{prior}, where we have defined the renormalized Planck mass $M_p \equiv m_p/H_0$. Furthermore, we have imposed priors on the parameter $K^2 M_p^4$ so as to ensure consistency with the approximation $K \ll 1$ underlying the analysis in Sec.~\ref{sec2}), this reduces to
\begin{equation}
\Delta \mathrm{AIC} = \chi_{\min}^{2}(\mathrm{KHDE}) - \chi_{\min}^{2}(\Lambda\mathrm{CDM}) + 4.
\end{equation}
According to Akaike’s scale, values of $\lvert \Delta \mathrm{AIC} \rvert < 2$ 
indicate that the two models are statistically equivalent. For 
$2 < \lvert \Delta \mathrm{AIC} \rvert < 6$, there is weak evidence in favor 
of the model with the smaller AIC value, while 
$6 < \lvert \Delta \mathrm{AIC} \rvert < 10$ corresponds to strong evidence. 
Finally, when $\lvert \Delta \mathrm{AIC} \rvert > 10$, there is decisive 
evidence supporting the model with the lower AIC.

At this stage, it is important to emphasize that, in order to avoid potential 
systematic biases in the comparison between the two models, the Hubble function 
for the $\Lambda$CDM model has also been derived numerically, following the same 
procedure adopted for our model. The set of free parameters for the 
$\Lambda$CDM model is $\{ H_{0}, \Omega_{m0}, r_{drag} \}$, while for the KHDE 
framework it is $\{ H_{0}, \Omega_{m0}, r_{drag}, K, c \}$. The priors adopted 
for the MCMC sampler are summarized in Table~\ref{prior}.


\subsection{Results}

We perform seven different constraints corresponding to the data set 
combinations listed in Table~\ref{dataset}. The best-fit parameters, together 
with the comparison to the $\Lambda$CDM model, are summarized in 
Table~\ref{bestfit}. In what follows, we present the remaining best-fit 
parameters for the seven data sets.

For the data set $\mathbf{D}_{1}$, which includes the OHD and BAO data, 
the best-fit parameters for the KHDE model are 
$H_{0} = 68.5_{-2.9}^{+2.0}$ (in units of $\mathrm{km \ s^{-1} \ Mpc^{-1}}$), 
$\Omega_{m0} = 0.270_{-0.015}^{+0.015}$, 
$r_{drag} = 147.2_{-3.4}^{+3.4}$ (in units of $\mathrm{Mpc}$), and 
$c = 0.98_{-0.54}^{+0.27}$, while the dimensionless parameter $K^{2}M_{p}^{4}$ can take all the values in the specific prior within the $1\sigma$.
The model does not provide a better fit than $\Lambda$CDM, since $\chi_{\min}^{2} - \chi_{\Lambda,\min}^{2} = +1.7$. According to Akaike’s scale, this corresponds to weak evidence in favor of the $\Lambda$CDM model, with $\Delta \mathrm{AIC} = +5.7$.

For the data set $\mathbf{D}_{2}$, which includes the PP and BAO data, 
the analysis of the MCMC chains yields the best-fit parameters 
$H_{0} = 69.1_{-5.4}^{+3.6}$, 
$\Omega_{m0} = 0.269_{-0.014}^{+0.014}$, 
$K^{2} M_{p}^{4} = 0.46_{-0.3}^{+0.3}$, 
$c = 1.06_{-0.24}^{+0.15}$, 
while $r_{drag}$ remains unconstrained within the adopted prior. 
In this case, the KHDE model provides a slightly better fit than 
$\Lambda$CDM, with $\chi_{\min}^{2} - \chi_{\Lambda,\min}^{2} = -0.8$. 
However, because of the larger number of free parameters, the AIC still indicates 
weak evidence in favor of $\Lambda$CDM, corresponding to 
$\Delta \mathrm{AIC} = +3.2$.

From data set $\mathbf{D}_{3}$, the MCMC analysis yields the cosmological
parameters $H_{0}=68.0_{-1.7}^{+1.7},~\Omega_{m0}=0.260_{-0.052}^{+0.030},
~c=1.21_{-0.48}^{+0.37}$, while both $K^{2}M_{p}^{4}$ and $r_{drag}$ remain
unconstrained. The $\Lambda$CDM model provides a better fit to this data set
than the KHDE, with the AIC indicating a weak preference in its favor.

The combination of all the late-time data as described by data set
$\mathbf{D}_{4}$, i.e. PP, BAO, and OHD data, leads to the best-fit parameters
$H_{0}=68.0_{-1.7}^{+1.7},~\Omega_{m0}=0.271_{-0.014}^{+0.014}$,
$r_{drag}=147.0_{-3.4}^{+3.4}$,  $K^{2} M_P^4 <0.603$ (in line with earlier results \cite{Hernandez-Almada:2021aiw,Hernandez-Almada:2021rjs}, thereby supporting the overall concordance of independent analyses) and $c=1.03_{-0.22}%
^{+0.15}$. The comparison of the statistical parameters gives $\mathbf{\chi
}_{\min}^{2}\mathbf{-\chi}_{\Lambda\min}^{2}=-0.6$, and $\Delta AIC=+3.4$,
from which we conclude that the KHDE fits the data better than the $\Lambda
$CDM, but the latter remains favored.

For the remaining three data sets, $\mathbf{D}_{5}$,~$\mathbf{D}_{6}$ and
$\mathbf{D}_{7}$, we replace the PP catalogue with the U3 supernova catalogue.
This replacement allows us to test the 
robustness of our constraints with a larger and more up-to-date data set.

For data set $\mathbf{D}_{5}$, the best-fit parameters are 
$H_{0}=68.3_{-5.4}^{+3.6}$, $\Omega_{m0}=0.270_{-0.015}^{+0.013}$, 
$c=1.34_{-0.35}^{+0.28}$, while both $K^{2}M_{p}^{4}$ and $r_{drag}$ remain 
unconstrained. We obtain $\chi_{\min}^{2}-\chi_{\Lambda,\min}^{2}=-2.8$ and 
$\Delta \mathrm{AIC}=+1.2$, indicating that although the KHDE provides a 
slightly better fit to this data set than $\Lambda$CDM, the two models are 
statistically equivalent.

From the analysis of the MCMC chains for data set $\mathbf{D}_{6}$, we obtain the 
best-fit parameters $H_{0}=66.8_{-1.8}^{+1.8}$, 
$\Omega_{m0}=0.273_{-0.037}^{+0.029}$, and $c>1.18$, while both $K^{2}M_{p}^{4}$ 
and $r_{drag}$ remain unconstrained. The comparison with $\Lambda$CDM yields 
$\chi_{\min}^{2}-\chi_{\Lambda,\min}^{2}=-1.1$ and 
$\Delta \mathrm{AIC}=+2.9$. Thus, although the KHDE model provides a slightly 
better fit to the data, the AIC indicates a weak preference for $\Lambda$CDM.

Finally, for data set $\mathbf{D}_{7}$ we obtain the best-fit parameters 
$H_{0}=67.1_{-1.8}^{+1.8}$, $\Omega_{m0}=0.271_{-0.014}^{+0.014}$, 
$r_{drag}=147.0_{-3.8}^{+3.4}$, and $c=1.31_{-0.37}^{+0.27}$, while 
$K^{2}M_{p}^{4}$ remains unconstrained. The comparison with $\Lambda$CDM gives 
$\chi_{\min}^{2}-\chi_{\Lambda,\min}^{2}=-2.6$ and $\Delta \mathrm{AIC}=1.4$, 
indicating that the two models are statistically equivalent.

We remark that the inclusion of the U3 catalogue provides stronger support for 
the KHDE model compared to the PP catalogue, and tends to favor larger values 
of the parameter $c$. Nevertheless, no significant tension is observed among 
the free parameters across the different data sets.

\begin{figure}[t]
\centering\includegraphics[width=0.9\textwidth]{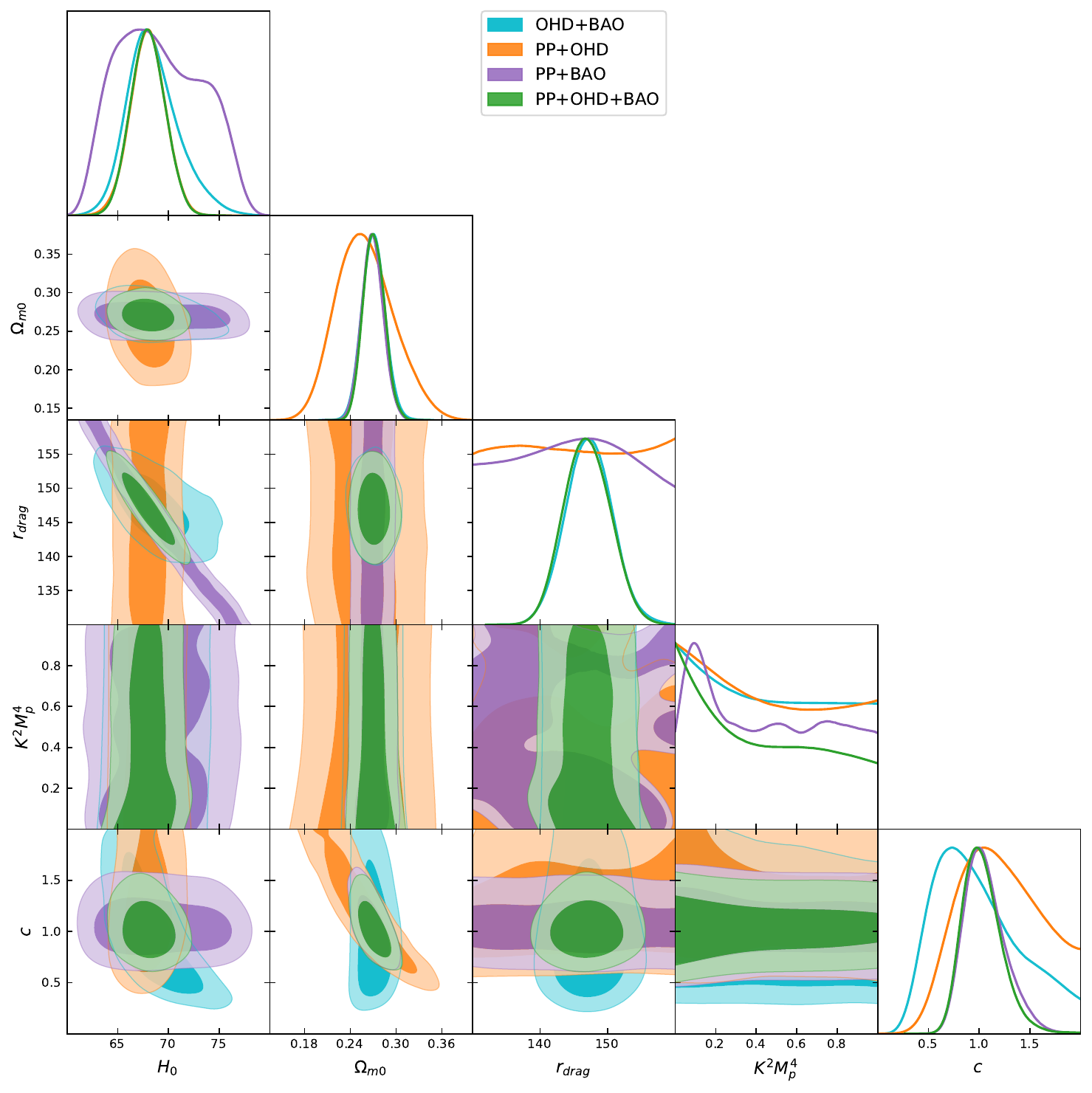}\caption{Confidence
space for the best-fit parameters for the KHDE model for the data sets
$\mathbf{D}_{1}$, $\mathbf{D}_{2}$,~$\mathbf{D}_{3}$ and $\mathbf{D}_{4}$}%
\label{fig1}%
\end{figure}

\begin{figure}[t]
\centering\includegraphics[width=0.9\textwidth]{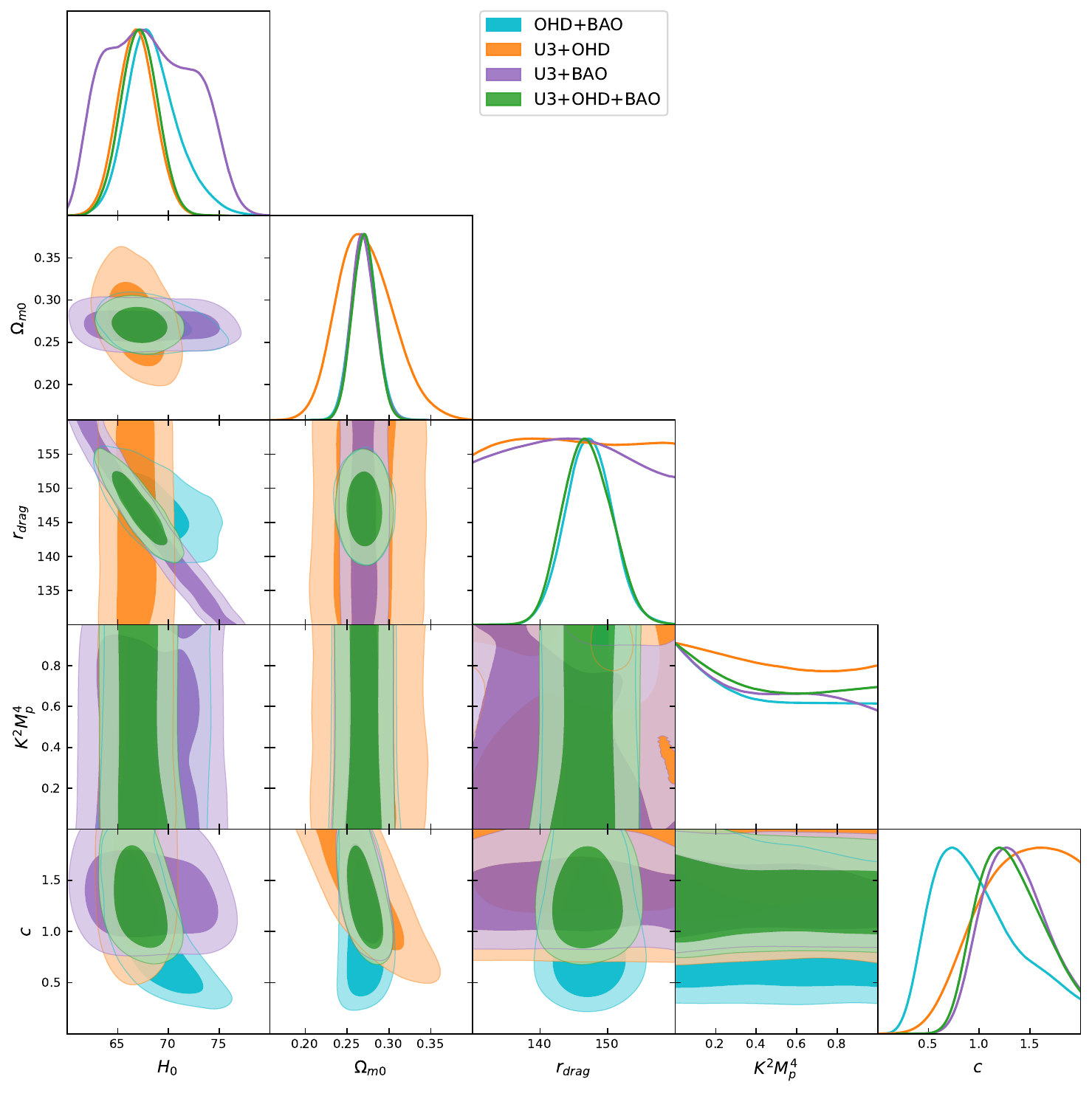}\caption{Confidence
space for the best-fit parameters for the KHDE model for the data sets
$\mathbf{D}_{1}$, $\mathbf{D}_{5}$,~$\mathbf{D}_{6}$ and $\mathbf{D}_{7}$}%
\label{fig2}%
\end{figure}

In Figs.~\ref{fig1} and \ref{fig2} we show the contour plots of the confidence 
regions for the best-fit parameters of the KHDE model. We find that the 
likelihood is maximized as ${K}^{2}{M}_{p}^{4}\rightarrow 0$, 
although the entire prior range remains consistent within the $1\sigma$ level. 
Therefore, with the exception of data set $\mathbf{D}_{2}$, none of the other 
data sets provide evidence for a significant deviation from standard holographic 
dark energy.

\begin{table}[tbp] \centering
\caption{Observational Constraints for the KHDE and the $\Lambda$CDM.}%
\begin{tabular}[c]{%
c@{\hspace{1em}}c@{\hspace{1em}}c@{\hspace{1em}}c@{\hspace{1em}}%
c@{\hspace{1em}}c@{\hspace{1em}}c@{\hspace{1em}}c}\hline\hline
& $\mathbf{H}_{0}$ & $\mathbf{\Omega}_{m0}$ & $\mathbf{r}_{drag}$ &
$\mathbf{K}^{2}\mathbf{M}_{p}^{4}$ & $\mathbf{c}$ & $\mathbf{\chi}_{\min}%
^{2}\mathbf{-\chi}_{\Lambda\min}^{2}$ & $\mathbf{AIC-AIC}_{\Lambda}$\\\hline
\multicolumn{8}{c}{\textbf{Data set }$\mathbf{D}_{1}$}\\
\textbf{KHDE} & $68.5_{-2.9}^{+2.0}$ & $0.270_{-0.015}^{+0.015}$ &
$147.2_{-3.4}^{+3.4}$ & $-$ & $0.98_{-0.54}^{+0.27}$ & $+1.7$ & $+5.7$\\
$\Lambda$\textbf{CDM} & $69.2_{-1.7}^{+1.7}$ & $0.296_{-0.012}^{+0.013}$ &
$147.2_{-3.1}^{+3.4}$ & $-$ & $-$ & $0$ & $0$\\
\multicolumn{8}{c}{\textbf{Data set }$\mathbf{D}_{2}$}\\
\textbf{KHDE} & $69.1_{-5.4}^{+3.6}$ & $0.269_{-0.014}^{+0.014}$ & $-$ &
$0.46_{-0.3}^{+0.3}$ & $1.06_{-0.24}^{+0.15}$ & $-0.8$ & $+3.2$\\
$\Lambda$\textbf{CDM} & $69.4_{-9.3}^{+3.2}$ & $0.309_{-0.012}^{+0.012}$ & $-$
& $-$ & $-$ & $0$ & $0$\\
\multicolumn{8}{c}{\textbf{Data set }$\mathbf{D}_{3}$}\\
\textbf{KHDE} & $68.0_{-1.7}^{+1.7}$ & $0.260_{-0.052}^{+0.030}$ & $-$ & $-$ &
$1.21_{-0.48}^{+0.37}$ & $+0.4$ & $+4.4$\\
$\Lambda$\textbf{CDM} & $67.6_{-1.7}^{+1.7}$ & $0.331_{-0.018}^{+0.018}$ & $-$
& $-$ & $-$ & $0$ & $0$\\
\multicolumn{8}{c}{\textbf{Data set }$\mathbf{D}_{4}$}\\
\textbf{KHDE} & $68.0_{-1.7}^{+1.7}$ & $0.271_{-0.014}^{+0.014}$ &
$147.0_{-3.4}^{+3.4}$ & $<0.603$ & $1.03_{-0.22}^{+0.15}$ & $-0.6$ & $+3.4$\\
$\Lambda$\textbf{CDM} & $68.5_{-1.6}^{+1.6}$ & $0.310_{-0.012}^{+0.010}$ &
$147.1_{-3.4}^{+3.4}$ & $-$ & $-$ & $0$ & $0$\\
\multicolumn{8}{c}{\textbf{Data set }$\mathbf{D}_{5}$}\\
\textbf{KHDE} & $68.3_{-5.4}^{+3.6}$ & $0.270_{-0.015}^{+0.013}$ & $-$ & $-$ &
$1.34_{-0.35}^{+0.28}$ & $-2.8$ & $+1.2$\\
$\Lambda$\textbf{CDM} & $-$ & $0.311_{-0.014}^{+0.014}$ & $-$ & $-$ & $-$ &
$0$ & $0$\\
\multicolumn{8}{c}{\textbf{Data set }$\mathbf{D}_{6}$}\\
\textbf{KHDE} & $66.8_{-1.8}^{+1.8}$ & $0.273_{-0.037}^{+0.029}$ & $-$ & $-$ &
$>1.18$ & $-1.1$ & $+2.9$\\
$\Lambda$\textbf{CDM} & $66.8_{-1.9}^{+1.9}$ & $0.351_{-0.025}^{+0.025}$ & $-$
& $-$ & $-$ & $0$ & $0$\\
\multicolumn{8}{c}{\textbf{Data set }$\mathbf{D}_{7}$}\\
\textbf{KHDE} & $67.1_{-1.8}^{+1.8}$ & $0.271_{-0.014}^{+0.014}$ &
$147.0_{-3.8}^{+3.4}$ & \thinspace$-$ & $1.31_{-0.37}^{+0.27}$ & $-2.6$ &
$+1.4$\\
$\Lambda$\textbf{CDM} & $68.6_{-1.7}^{+1.7}$ & $0.311_{-0.014}^{+0.013}$ &
$146.8_{-3.4}^{+3.4}$ & $-$ & $-$ & $0$ & $0$\\\hline\hline
\end{tabular}
\label{bestfit}%
\end{table}%

\section{Conclusions}
\label{sec4}

Holographic Dark Energy (HDE) is a well-established framework inspired by the holographic principle of quantum gravity, where the dark energy density is assumed to scale inversely with the square of a characteristic infrared (IR) cutoff, typically identified with the future event horizon. To incorporate potential departures from standard thermodynamic behavior in high-energy or quantum gravity regimes, several extensions of HDE have been introduced that rely on generalized entropy formalisms. In this context, 
Kaniadakis entropy arises as a deformation of the standard Boltzmann--Gibbs entropy, motivated by relativistic statistical mechanics. It is characterized by a dimensionless parameter $-1<K<1$, which quantifies the deviation from extensivity, with $K \to 0$ restoring the usual Bekenstein--Hawking entropy. When this generalized entropy is implemented in the HDE framework, one obtains the Kaniadakis Holographic Dark Energy (KHDE) scenario, featuring modified Friedmann equations and a phenomenology richer than in the standard case. In contrast to other generalized entropy approaches, such as Tsallis or Barrow entropy, the Kaniadakis formalism introduces deformations with a solid foundation in relativistic kinetic theory, thereby offering a conceptually distinct path toward exploring possible extensions of Einstein’s theory in the context of dark energy dynamics.

In this work, we have tested the KHDE scenario against a multiple late-time cosmological probes, including Type Ia supernovae from PP and U3, CC measurements of the Hubble parameter and BAO data from DESI DR2. Our results show that, while KHDE can accommodate different data combinations and in some cases achieves a marginally better fit than $\Lambda$CDM, the standard cosmological model remains slightly statistically favored according to the Akaike Information Criterion. Nevertheless, these findings underscore the relevance of KHDE as a competitive alternative, offering meaningful insights into the role of generalized entropy in the dynamics of dark energy. In particular, they highlight the model’s ability to reproduce current observations with notable accuracy, while at the same time opening up new theoretical perspectives that link thermodynamical principles with cosmological evolution.

Further aspects need to be investigated: first, it would be valuable to extend the present analysis by examining KHDE at the perturbative level, with particular emphasis on structure formation and growth. A confrontation with high-precision cosmological probes - such as Cosmic Microwave Background (CMB) temperature and polarization spectra, weak gravitational lensing surveys, and measurements related to the $\sigma_8$ parameter - could provide a stringent test of their viability. Incorporating these observables would not only sharpen the constraints on the underlying parameters but also clarify whether generalized entropy frameworks can successfully capture both background expansion and perturbative dynamics. Such a holistic approach is expected to shed light on the broader question of whether extended entropic descriptions are best interpreted within the holographic dark energy paradigm or as manifestations of more radical proposals connecting gravity and thermodynamics. A comprehensive investigation along these lines will be the focus of a forthcoming study.

\begin{acknowledgments}
The research of GGL is supported by the postdoctoral fellowship program of the
University of Lleida. GGL gratefully acknowledges the contribution of the LISA
Cosmology Working Group (CosWG), as well as support from the COST Actions
CA21136 - \textit{Addressing observational tensions in cosmology with
systematics and fundamental physics (CosmoVerse)} - CA23130, \textit{Bridging
high and low energies in search of quantum gravity (BridgeQG)} and CA21106 -
\textit{COSMIC WISPers in the Dark Universe: Theory, astrophysics and
experiments (CosmicWISPers)}. AP thanks the support of VRIDT through
Resoluci\'{o}n VRIDT No. 096/2022 and Resoluci\'{o}n VRIDT No. 098/2022. Part
of this study was supported by FONDECYT 1240514.
\end{acknowledgments}

\bibliography{Bib}

\end{document}